\documentclass{article}
\usepackage{spconf}
\usepackage{makecell}
\usepackage{arydshln}

\usepackage{amsmath,bm,amssymb,amsthm,amsfonts}

\usepackage{listings}%
\usepackage[font=small,labelfont=bf]{caption}
\usepackage{subcaption}
\usepackage{cite}
\usepackage{enumitem} \setlist[itemize]{leftmargin=*}

\usepackage{algorithmic}
\usepackage[ruled,linesnumbered]{algorithm2e}

\usepackage{graphicx}%
\usepackage{booktabs}
\usepackage{setspace,hyperref}
\usepackage{color,xcolor}
\usepackage{etoolbox}
\usepackage{textcomp}
\usepackage{amsmath}
\usepackage{circuitikz}
\usepackage{tikz}
\usetikzlibrary{positioning}
\usetikzlibrary{shapes.symbols}
\usetikzlibrary{angles,quotes}
\usepackage{pgfkeys}

\newcommand{\RNum}[1]{\uppercase\expandafter{\romannumeral #1\relax}}

\input{irs.tex}
\input{basestation.tex}

\makeatletter

\newif\ifleftantennas

\pgfkeys{
  /mimonode/leftantennas/.is if=leftantennas,
  /pgf/.cd,
  right antennas/.code={\pgfkeys{/mimonode/leftantennas=false}},
  left antennas/.code={\pgfkeys{/mimonode/leftantennas=true}},
  circle fill color/.initial=white,
  circle stroke color/.initial=black,
  circle radius/.initial=2mm,
  antenna offset/.initial=0.3cm,
  antenna yshift/.initial=0.cm,
  antenna base height/.initial=0.2 cm,
  antenna side/.initial=0.4cm,
}

\def\mimoshapesdrawantenna{
  \pgf@xa=\pgf@x 
  \pgf@ya=\pgf@y
  \pgfpathmoveto{\pgfpoint{\pgf@x}{\pgf@y}}

  \advance\pgf@xa by \antennaoffset
  \pgfpathlineto{\pgfpoint{\pgf@xa}{\pgf@ya}}
  \pgfpathclose
  \pgfmoveto{\pgfpoint{\pgf@xa}{\pgf@ya}}

  \pgf@xb=\pgf@xa \pgf@yb=\pgf@ya
  \advance\pgf@yb by \antennabaseheight cm
  \pgfpathlineto{\pgfpoint{\pgf@xb}{\pgf@yb}}
  
  \pgf@process{
    \pgfpointadd{\pgfpoint{\pgf@xb}{\pgf@yb}}{\pgfpointpolar{60}{\pgfkeysvalueof{/pgf/antenna side}}}
  }
  \pgf@xc=\pgf@x 
  \pgf@yc=\pgf@y
  \pgfpathlineto{\pgfpoint{\pgf@x}{\pgf@y}}

  \pgf@process{
    \pgfpointadd{\pgfpoint{\pgf@xc}{\pgf@yc}}{\pgfpointpolar{180}{\pgfkeysvalueof{/pgf/antenna side}}}
  }
  \pgfpathlineto{\pgfpoint{\pgf@x}{\pgf@y}}
  
  \pgf@process{
    \pgfpointadd{\pgfpoint{\pgf@xb}{\pgf@yb}}{\pgfpointpolar{240}{\pgfkeysvalueof{/pgf/antenna side}}}
  }
  \pgfpathlineto{\pgfpoint{\pgf@xb}{\pgf@yb}}
}

\def\drawellipses{
  \pgf@xa=\pgf@x 
  \advance\pgf@xa by \antennaoffset
  \pgf@ya=\pgf@y
  
  \pgf@process{\pgfpathcircle{\pgfpoint{\pgf@xa}{\pgf@ya+\antennabaseheight-1pt}}{1pt}}
  \pgf@process{\pgfpathcircle{\pgfpoint{\pgf@xa}{\pgf@ya+\antennabaseheight+2pt}}{1pt}}
  \pgf@process{\pgfpathcircle{\pgfpoint{\pgf@xa}{\pgf@ya+\antennabaseheight+5pt}}{1pt}}
}

\def\antennaStartXCoordinate{
  \ifleftantennas
  \pgf@x=-.5\wd\pgfnodeparttextbox
  \pgfmathsetlength\pgf@xc{-\pgfkeysvalueof{/pgf/inner xsep}}
  \advance\pgf@x by \pgf@xc%
  \setlength{\pgf@xa}{-\pgfshapeminwidth}
  \ifdim\pgf@x>.5\pgf@xa
  \pgf@x=.5\pgf@xa
  \fi
  \else
  \pgf@x=.5\wd\pgfnodeparttextbox 
  \pgfmathsetlength\pgf@xc{\pgfkeysvalueof{/pgf/inner xsep}}
  \advance\pgf@x by \pgf@xc%
  \setlength{\pgf@xa}{\pgfshapeminwidth}
  \ifdim\pgf@x<.5\pgf@xa
  \pgf@x=.5\pgf@xa
  \fi
  \fi
}

\pgfdeclareshape{mimoone}{

  \savedmacro\antennaoffset{
    \ifleftantennas
      \def\antennaoffset{-\pgfkeysvalueof{/pgf/antenna offset}}
    \else
      \def\antennaoffset{\pgfkeysvalueof{/pgf/antenna offset}}
    \fi
  }

  \savedmacro\antennabaseheight{
    \def\antennabaseheight{\pgfkeysvalueof{/pgf/antenna base height}}
  }

  \savedmacro\antennayshift{
    \def\antennayshift{\pgfkeysvalueof{/pgf/antenna yshift}}
  }

  \savedmacro\shapewidth{
    \def\shapewidth{\pgfkeysvalueof{/pgf/minimum width}}
  }

  \savedmacro\shapeheight{
    \def\shapeheight{\pgfkeysvalueof{/pgf/minimum height}}
  }

  \savedanchor\northeast{%
    \pgf@y=.5\ht\pgfnodeparttextbox 
    \pgf@x=.5\wd\pgfnodeparttextbox 
    \pgfmathsetlength\pgf@xc{\pgfkeysvalueof{/pgf/inner xsep}}
    \advance\pgf@x by \pgf@xc%
    \pgfmathsetlength\pgf@yc{\pgfkeysvalueof{/pgf/inner ysep}}
    \advance\pgf@y by \pgf@yc%
    \setlength{\pgf@xa}{\shapewidth}
    \ifdim\pgf@x<.5\pgf@xa
    \pgf@x=.5\pgf@xa
    \fi
    \setlength{\pgf@ya}{\shapeheight}
    \ifdim\pgf@y<.5\pgf@ya
    \pgf@y=.5\pgf@ya
    \fi
    \pgfmathsetlength\pgf@xa{\pgfkeysvalueof{/pgf/outer xsep}}%
    \advance\pgf@x by\pgf@xa%
    \pgfmathsetlength\pgf@ya{\pgfkeysvalueof{/pgf/outer ysep}}%
    \advance\pgf@y by\pgf@ya%
  }

  \savedanchor\southwest{%
    \pgf@y=-.5\ht\pgfnodeparttextbox 
    \pgf@x=-.5\wd\pgfnodeparttextbox 
    \pgfmathsetlength\pgf@xc{-\pgfkeysvalueof{/pgf/inner xsep}}
    \advance\pgf@x by \pgf@xc%
    \pgfmathsetlength\pgf@yc{-\pgfkeysvalueof{/pgf/inner ysep}}
    \advance\pgf@y by \pgf@yc%
    \setlength{\pgf@xa}{-\shapewidth}
    \ifdim\pgf@x>.5\pgf@xa
    \pgf@x=.5\pgf@xa
    \fi
    \setlength{\pgf@ya}{-\shapeheight}
    \ifdim\pgf@y>.5\pgf@ya
    \pgf@y=.5\pgf@ya
    \fi
    \pgfmathsetlength\pgf@xa{-\pgfkeysvalueof{/pgf/outer xsep}}%
    \advance\pgf@x by\pgf@xa%
    \pgfmathsetlength\pgf@ya{-\pgfkeysvalueof{/pgf/outer ysep}}%
    \advance\pgf@y by\pgf@ya%
  }

  \savedanchor{\anchorA}{
    \pgf@y=.5\ht\pgfnodeparttextbox
    \setlength{\pgf@ya}{\pgfshapeminheight}
    \ifdim\pgf@y<.5\pgf@ya
    \pgf@y=.5\pgf@ya
    \fi
    \pgf@y=0.\pgf@y
    \advance\pgf@y by \antennayshift
    \antennaStartXCoordinate
  }

  \inheritanchorborder[from=rectangle]
  
  \inheritanchor[from=rectangle]{center}
  \inheritanchor[from=rectangle]{north}
  \inheritanchor[from=rectangle]{east}
  \inheritanchor[from=rectangle]{south}
  \inheritanchor[from=rectangle]{west}
  \inheritanchor[from=rectangle]{north east}
  \inheritanchor[from=rectangle]{north west}
  \inheritanchor[from=rectangle]{south west}
  \inheritanchor[from=rectangle]{south east}

  \anchor{A}{
    \pgf@process{\anchorA}
  }

  \anchor{text}
  {
    \pgf@process{\pgfpointorigin}
    \advance\pgf@x by -.5\wd\pgfnodeparttextbox%
    \advance\pgf@y by -.5\ht\pgfnodeparttextbox%
    \advance\pgf@y by +.5\dp\pgfnodeparttextbox%
  }
  
  \anchor{first antenna base start}{
    \pgf@process{\anchorA}
  }

  \anchor{first antenna base end}{
    \pgf@process{\anchorA}
    \advance\pgf@x by \antennaoffset%
    \advance\pgf@y by 0.2cm%
  }

  \anchor{first antenna}{
    \pgf@process{\anchorA}
    \advance\pgf@x by \antennaoffset%
    \advance\pgf@y by 0.4cm%
  }

  \backgroundpath{
    \pgfpathrectanglecorners
    { 
      \pgfpointadd{\southwest}{\pgfpoint{\pgfkeysvalueof{/pgf/outer xsep}}{\pgfkeysvalueof{/pgf/outer ysep}}}
    }
    { 
      \pgfpointadd{\northeast}{\pgfpointscale{-1}{\pgfpoint{\pgfkeysvalueof{/pgf/outer xsep}}{\pgfkeysvalueof{/pgf/outer ysep}}}}
    }

    \pgf@process{\anchorA}
    \mimoshapesdrawantenna
  }
}

\pgfdeclareshape{mimotwo}{

  \inheritsavedanchors[from=mimoone]  

  \savedanchor{\anchorA}{
    \pgf@y=.5\ht\pgfnodeparttextbox
    \setlength{\pgf@ya}{\pgfshapeminheight}
    \ifdim\pgf@y<.5\pgf@ya
    \pgf@y=.5\pgf@ya
    \fi
    \pgf@y=0.4\pgf@y
    \advance\pgf@y by \antennayshift
    \antennaStartXCoordinate
  }

  \savedanchor{\anchorB}{
    \pgf@y=-.5\ht\pgfnodeparttextbox
    \setlength{\pgf@ya}{-\pgfshapeminheight}
    \ifdim\pgf@y>.5\pgf@ya
    \pgf@y=.5\pgf@ya
    \fi
    \pgf@y=0.6\pgf@y
    \advance\pgf@y by \antennayshift
    \antennaStartXCoordinate
  }

  \inheritanchorborder[from=mimoone]
  
  \inheritanchor[from=mimoone]{center}
  \inheritanchor[from=mimoone]{north}
  \inheritanchor[from=mimoone]{east}
  \inheritanchor[from=mimoone]{south}
  \inheritanchor[from=mimoone]{west}
  \inheritanchor[from=mimoone]{north east}
  \inheritanchor[from=mimoone]{north west}
  \inheritanchor[from=mimoone]{south west}
  \inheritanchor[from=mimoone]{south east}
  \inheritanchor[from=mimoone]{text}
  \inheritanchor[from=mimoone]{A}
  \inheritanchor[from=mimoone]{first antenna base start}
  \inheritanchor[from=mimoone]{first antenna base end}
  \inheritanchor[from=mimoone]{first antenna}
  
  \anchor{B}{
    \pgf@process{\anchorB}
  }

  \anchor{second antenna base start}{
    \pgf@process{\anchorB}
  }

  \anchor{second antenna base end}{
    \pgf@process{\anchorB}
    \advance\pgf@x by \antennaoffset%
    \advance\pgf@y by 0.2cm%
  }

  \anchor{second antenna}{
    \pgf@process{\anchorB}
    \advance\pgf@x by \antennaoffset%
    \advance\pgf@y by 0.4cm%
  }

  \backgroundpath{
    \pgfpathrectanglecorners
    { 
      \pgfpointadd{\southwest}{\pgfpoint{\pgfkeysvalueof{/pgf/outer xsep}}{\pgfkeysvalueof{/pgf/outer ysep}}}
    }
    { 
      \pgfpointadd{\northeast}{\pgfpointscale{-1}{\pgfpoint{\pgfkeysvalueof{/pgf/outer xsep}}{\pgfkeysvalueof{/pgf/outer ysep}}}}
    }

    \pgf@process{\anchorA}
    \mimoshapesdrawantenna

    \pgf@process{\anchorB}
    \mimoshapesdrawantenna
  }
}

\pgfdeclareshape{mimothree}{

  \inheritsavedanchors[from=mimotwo]

  \savedanchor{\anchorA}{
    \pgf@y=.5\ht\pgfnodeparttextbox
    \setlength{\pgf@ya}{\pgfshapeminheight}
    \ifdim\pgf@y<.5\pgf@ya
    \pgf@y=.5\pgf@ya
    \fi
    \pgf@y=0.5\pgf@y
    \advance\pgf@y by \antennayshift
    \antennaStartXCoordinate
  }

  \savedanchor{\anchorC}{
    \pgf@y=-.5\ht\pgfnodeparttextbox
    \setlength{\pgf@ya}{-\pgfshapeminheight}
    \ifdim\pgf@y>.5\pgf@ya
    \pgf@y=.5\pgf@ya
    \fi
    \pgf@y=0.9\pgf@y
    \advance\pgf@y by \antennayshift
    \antennaStartXCoordinate
  }

  \savedanchor{\anchorB}{
    \pgf@y=.5\ht\pgfnodeparttextbox
    \setlength{\pgf@ya}{\pgfshapeminheight}
    \ifdim\pgf@y<.5\pgf@ya
    \pgf@y=.5\pgf@ya
    \fi
    \pgf@ya=0.5\pgf@y
    \pgf@y=-.5\ht\pgfnodeparttextbox
    \setlength{\pgf@yc}{-\pgfshapeminheight}
    \ifdim\pgf@y>.5\pgf@yc
    \pgf@y=.5\pgf@yc
    \fi
    \pgf@yc=0.9\pgf@y
    \pgf@y=0.5\pgf@yc
    \advance\pgf@y by 0.5\pgf@ya
    \advance\pgf@y by \antennayshift
    \antennaStartXCoordinate
  }

  \inheritanchorborder[from=mimotwo]

  \inheritanchor[from=mimotwo]{center}
  \inheritanchor[from=mimotwo]{north}
  \inheritanchor[from=mimotwo]{east}
  \inheritanchor[from=mimotwo]{south}
  \inheritanchor[from=mimotwo]{west}
  \inheritanchor[from=mimotwo]{north east}
  \inheritanchor[from=mimotwo]{north west}
  \inheritanchor[from=mimotwo]{south west}
  \inheritanchor[from=mimotwo]{south east}
  \inheritanchor[from=mimotwo]{text}
  \inheritanchor[from=mimotwo]{A}
  \inheritanchor[from=mimotwo]{B}
  \inheritanchor[from=mimotwo]{first antenna base start}
  \inheritanchor[from=mimotwo]{second antenna base start}
  \inheritanchor[from=mimotwo]{first antenna base end}
  \inheritanchor[from=mimotwo]{second antenna base end}
  \inheritanchor[from=mimotwo]{first antenna}
  \inheritanchor[from=mimotwo]{second antenna}
  
  \anchor{C}{\anchorC}

  \anchor{third antenna base start}{
    \pgf@process{\anchorC}
  }
  \anchor{third antenna base end}{
    \pgf@process{\anchorC}
    \advance\pgf@x by \antennaoffset%
    \advance\pgf@y by 0.2cm%
  }
  \anchor{third antenna}{
    \pgf@process{\anchorC}
    \advance\pgf@x by \antennaoffset%
    \advance\pgf@y by 0.4cm%
  }

  \backgroundpath{
    \pgfpathrectanglecorners
    { 
      \pgfpointadd{\southwest}{\pgfpoint{\pgfkeysvalueof{/pgf/outer xsep}}{\pgfkeysvalueof{/pgf/outer ysep}}}
    }
    { 
      \pgfpointadd{\northeast}{\pgfpointscale{-1}{\pgfpoint{\pgfkeysvalueof{/pgf/outer xsep}}{\pgfkeysvalueof{/pgf/outer ysep}}}}
    }

    \pgf@process{\anchorA}
    \mimoshapesdrawantenna

    \pgf@process{\anchorB}
    \mimoshapesdrawantenna

    \pgf@process{\anchorC}
    \mimoshapesdrawantenna
  }
}

\pgfdeclareshape{mimoind}{

  \inheritsavedanchors[from=mimothree]

  \inheritanchorborder[from=mimothree]

  \inheritanchor[from=mimothree]{center}
  \inheritanchor[from=mimothree]{north}
  \inheritanchor[from=mimothree]{east}
  \inheritanchor[from=mimothree]{south}
  \inheritanchor[from=mimothree]{west}
  \inheritanchor[from=mimothree]{north east}
  \inheritanchor[from=mimothree]{north west}
  \inheritanchor[from=mimothree]{south west}
  \inheritanchor[from=mimothree]{south east}
  \inheritanchor[from=mimothree]{text}
  \inheritanchor[from=mimothree]{A}
  \inheritanchor[from=mimothree]{B}
  \inheritanchor[from=mimothree]{C}
  \inheritanchor[from=mimothree]{first antenna base start}
  \inheritanchor[from=mimothree]{second antenna base start}
  \inheritanchor[from=mimothree]{third antenna base start}
  \inheritanchor[from=mimothree]{first antenna base end}
  \inheritanchor[from=mimothree]{second antenna base end}
  \inheritanchor[from=mimothree]{third antenna base end}
  \inheritanchor[from=mimothree]{first antenna}
  \inheritanchor[from=mimothree]{second antenna}
  \inheritanchor[from=mimothree]{third antenna}

  \backgroundpath{
    \pgfpathrectanglecorners
    { 
      \pgfpointadd{\southwest}{\pgfpoint{\pgfkeysvalueof{/pgf/outer xsep}}{\pgfkeysvalueof{/pgf/outer ysep}}}
    }
    { 
      \pgfpointadd{\northeast}{\pgfpointscale{-1}{\pgfpoint{\pgfkeysvalueof{/pgf/outer xsep}}{\pgfkeysvalueof{/pgf/outer ysep}}}}
    }

    \pgf@process{\anchorA}
    \mimoshapesdrawantenna

    \pgf@process{\anchorB}
    \drawellipses
    
    \pgf@process{\anchorC}
    \mimoshapesdrawantenna
  }
}

\pgfdeclareshape{boxtwo}{

  \inheritsavedanchors[from=mimoone]

  \inheritanchorborder[from=mimoone]

  \inheritanchor[from=mimotwo]{center}
  \inheritanchor[from=mimotwo]{north}
  \inheritanchor[from=mimotwo]{east}
  \inheritanchor[from=mimotwo]{south}
  \inheritanchor[from=mimotwo]{west}
  \inheritanchor[from=mimotwo]{north east}
  \inheritanchor[from=mimotwo]{north west}
  \inheritanchor[from=mimotwo]{south west}
  \inheritanchor[from=mimotwo]{south east}
  \inheritanchor[from=mimotwo]{text}

  \anchor{A}{
    \pgf@process{\northeast}%
    \pgf@x=0pt
    \pgf@y=0.5\pgf@y%
  }

  \anchor{B}{
    \pgf@process{\northeast}%
    \pgf@x=0pt
    \pgf@y=-0.5\pgf@y%
  }

  \backgroundpath{
    {
      \pgfpathrectanglecorners
      { 
        \pgfpointadd{\southwest}{\pgfpoint{\pgfkeysvalueof{/pgf/outer xsep}}{\pgfkeysvalueof{/pgf/outer ysep}}}
      }
      { 
        \pgfpointadd{\northeast}{\pgfpointscale{-1}{\pgfpoint{\pgfkeysvalueof{/pgf/outer xsep}}{\pgfkeysvalueof{/pgf/outer ysep}}}}
      }
      \pgfusepath{stroke,fill}
    }
    
    {
      \csname pgf@anchor@\pgf@sm@shape@name @A\endcsname
      \pgfcircle{\pgfpoint{\pgf@x}\pgf@y{}}{\pgfkeysvalueof{/pgf/circle radius}}
      \csname pgf@anchor@\pgf@sm@shape@name @B\endcsname
      \pgfcircle{\pgfpoint{\pgf@x}\pgf@y{}}{\pgfkeysvalueof{/pgf/circle radius}}
      \pgfsetfillcolor{\pgfkeysvalueof{/pgf/circle fill color}}
      \pgfsetstrokecolor{\pgfkeysvalueof{/pgf/circle stroke color}}
      \pgfusepath{stroke,fill}
    }
  }
}

\pgfdeclareshape{boxthree}{

  \inheritsavedanchors[from=boxtwo]

  \inheritanchorborder[from=boxtwo]

  \inheritanchor[from=boxtwo]{center}
  \inheritanchor[from=boxtwo]{north}
  \inheritanchor[from=boxtwo]{east}
  \inheritanchor[from=boxtwo]{south}
  \inheritanchor[from=boxtwo]{west}
  \inheritanchor[from=boxtwo]{north east}
  \inheritanchor[from=boxtwo]{north west}
  \inheritanchor[from=boxtwo]{south west}
  \inheritanchor[from=boxtwo]{south east}
  \inheritanchor[from=boxtwo]{text}

  \anchor{A}{
    \northeast
    \pgf@x=0pt
    \pgf@y=0.6\pgf@y%
  }

  \anchor{B}{
    \pgf@x=0pt
    \pgf@y=0pt
  }

  \anchor{C}{
    \northeast
    \pgf@x=0pt
    \pgf@y=-0.6\pgf@y%
  }

  \backgroundpath{
    {
      \pgfpathrectanglecorners
      { 
        \pgfpointadd{\southwest}{\pgfpoint{\pgfkeysvalueof{/pgf/outer xsep}}{\pgfkeysvalueof{/pgf/outer ysep}}}
      }
      { 
        \pgfpointadd{\northeast}{\pgfpointscale{-1}{\pgfpoint{\pgfkeysvalueof{/pgf/outer xsep}}{\pgfkeysvalueof{/pgf/outer ysep}}}}
      }
      \pgfusepath{stroke,fill}
    }
    
    {
      \csname pgf@anchor@\pgf@sm@shape@name @A\endcsname
      \pgfcircle{\pgfpoint{\pgf@x}\pgf@y{}}{\pgfkeysvalueof{/pgf/circle radius}}
      \csname pgf@anchor@\pgf@sm@shape@name @B\endcsname
      \pgfcircle{\pgfpoint{\pgf@x}\pgf@y{}}{\pgfkeysvalueof{/pgf/circle radius}}
      \csname pgf@anchor@\pgf@sm@shape@name @C\endcsname
      \pgfcircle{\pgfpoint{\pgf@x}\pgf@y{}}{\pgfkeysvalueof{/pgf/circle radius}}
      \pgfsetfillcolor{\pgfkeysvalueof{/pgf/circle fill color}}
      \pgfsetstrokecolor{\pgfkeysvalueof{/pgf/circle stroke color}}
      \pgfusepath{stroke,fill}
      }
  }
}

\makeatother

\DeclareMathOperator{\diag}{diag}
\DeclareMathOperator{\supp}{supp}

\DeclareMathOperator{\vect}{vec}
\title{Structure-aware Sparse Bayesian Learning-based Channel Estimation for Intelligent Reflecting Surface-aided MIMO}
\name{Yanbin He and Geethu Joseph}
\address{Department of Microelectronics, Delft University of Technology, Delft, The Netherlands\\Emails: \{y.he-1, g.joseph\}@tudelft.nl.}
\begin{document}
%
\setlength{\belowdisplayskip}{4pt}
\setlength{\abovedisplayskip}{3pt} 
\maketitle
\begin{abstract}
This paper presents novel cascaded channel estimation techniques for an intelligent reflecting surface-aided multiple-input multiple-output system. Motivated by the channel angular sparsity at higher frequency bands, the channel estimation problem is formulated as a sparse vector recovery problem with an inherent Kronecker structure. We solve the problem using the sparse Bayesian learning framework which leads to a non-convex optimization problem. We offer two solution techniques to the problem based on alternating minimization and singular value decomposition. Our simulation results illustrate the superior performance of our methods in terms of accuracy and run time compared with the existing works.
\end{abstract}
%
\section{Introduction}
\label{sec:intro}

An intelligent reflecting surface (IRS) is a digitally controlled meta-surface containing a large number of passive reflecting elements. By reconfiguring the reflection coefficient of each element, IRS controls the wireless channel to improve the coverage and capacity of the communication system \cite{wu2021intelligent,wei2021channel,bjornson2022reconfigurable}. However, to enhance the channel properties via IRS, obtaining accurate channel state information is inevitable. Therefore, in this paper, we address the uplink channel estimation problem for an IRS-aided multi-input multi-output (MIMO) system by exploiting the intrinsic channel structure.

\noindent\textbf{Related works:}  
Early works on channel estimation for IRS-aided communication systems focused on unstructured channel models \cite{swindlehurst2022channel}, employing least squares or linear minimum mean square error estimators \cite{zheng2022survey}. However, in higher frequency bands (for example, millimeter wave or terahertz band) both mobile station (MS)-IRS and IRS-base stations (BS) channels exhibit strong sparsity in the angular domain \cite{zheng2022survey}. This observation motivated the IRS-aided channel estimation algorithms to explore the intrinsic sparsity of the channel,
reducing the pilot overhead \cite{zheng2022survey}.
Recent estimators further enhanced the accuracy by accounting for additional structures along with sparsity. Some examples are clustered sparsity structure in the angular domain~\cite{you2022structured} and joint sparsity in a multiuser setting~\cite{wei2021channel,zhou2022channel}.
Most studies use orthogonal matching pursuit (OMP)-based methods. Despite low complexity, their heuristic nature leads to inferior channel estimation accuracy compared to other sparsity-driven approaches.
An alternative approach is the iterative reweighted method-based sparse channel estimation~\cite{he2020channel}, but it
does not incorporate any additional signal structure.
\cite{xu2022sparse} presents a sparse Bayesian learning (SBL) scheme 
to handle the inherent Kronecker structure of the cascaded BS-IRS-MS sparse channel. However, the derivation of the SBL algorithm relied on several approximations leading to a suboptimal estimation accuracy~\cite{chang2021sparse}. 
Hence, we seek a novel channel estimator that exploits the Kronecker-sparse structure of the cascaded channel and offers improved estimation accuracy and complexity.


\noindent\textbf{Contributions:} Our contributions are two novel SBL channel estimation algorithms for an IRS-aided system:
\begin{itemize}
    \item \emph{Alternating minimization (AM)-based:} This method solves the underlying optimization problem of the SBL algorithm exactly using the AM procedure, inheriting the convergence property of the SBL algorithm.
    \item \emph{Singular value decomposition (SVD)-based:} The second method uses a simple approximation to obtain the SBL algorithm. However, the resulting algorithm is faster and more accurate than the state-of-the-art Kronecker-SBL~\cite{xu2022sparse}.
\end{itemize}

Overall, we derive two SBL-based channel estimators that exploit the Kronecker-sparse structure, leading to improved pilot overhead. The algorithms can be of independent interest because Kronecker-sparse structure naturally arises in a basis expansion problem with multiple unknown parameters.

\section{Cascaded Channel Estimation Problem}
\label{sec:channelmodel}

\begin{figure}[t]
\centering
\resizebox{180pt}{106pt}{%
\begin{tikzpicture}[every node/.append style={scale=1}]

  \path (0,0) coordinate (origin);
  \node[IRS, element fill color=green!30, fill=red!30, minimum size=2.5cm] at (2.5,-3) (irs) {};
  \draw[] (3,-2.7) node[scale=1,rotate=0] {IRS};
  \draw[] (4.04,-2.1) node[scale=1,rotate=0] {$\cdots$};
  

  \path (-0.5,1) coordinate (mimoindposl);
  \path (7,1) coordinate (mimoindposr);


  \tikzset{test mimo node/.style={fill=blue!30,draw,thick,
      antenna offset=0.3cm, 
      minimum width=0.5cm,
      minimum height=2cm, 
      antenna base height=0.2cm,
      antenna side=0.4cm,
      antenna yshift=0.3cm
    }}
  \def\txrxsep{4cm}

  \draw (mimoindposl)  node[draw,test mimo node,mimoind] (mimoindtxnode) {BS};
  \draw (mimoindposr)  node[draw,test mimo node,mimoind,left antennas] (mimoindrxnode) {MS};

  \node[] (1) at (0.35,1.5)  {};
  \node[] (1B) at (0.35,1)  {};
  \node[] (12) at (1.8,-0.6)  {};  
  \node[] (121) at (0.35,0.6)  {};  
  \node[] (122) at (0.35,1.3)  {};  
  \node[] (123) at (0.35,2.0)  {};  
  
  \node[] (2) at (2.5,-1.8)  {};
  \node[] (2B) at (2,-1.8)  {};
  \node[] (2C) at (4,-1.8)  {};
  
  \node[] (23) at (4.9,-0.4)  {};
  \node[] (231) at (3.5,-1.8)  {};
  \node[] (232) at (3,-1.8)  {};
  \node[] (233) at (2.5,-1.8)  {};  
  
  \node[] (3) at (6.15,2)  {};
  \node[] (3B) at (6.15,1)  {};
  
  \draw[->] (23)--(231);
  \draw[->] (23)--(232);
  \draw[->] (23)--(233);
  
  \draw[->] (3)--(23);
  
  \draw[->] (2)--(12);
  \draw[->] (12)--(121);
  \draw[->] (12)--(122);
  \draw[->] (12)--(123);
  
  \draw[thick,black,->] (0.35,2.0) -- (0.35,0.5) node[black,right] {};
  \draw[thick,black,<->] (1.2,-1.8) -- (5,-1.8) node[black,right] {};
  \draw[thick,black,->] (6.13,2) -- (6.13,0.5) node[black,right] {};
  
  \draw pic[draw,angle radius=0.8cm,"$\alpha_{\mathrm{BS}}$",pic text options={shift={(0.5,0)}}] {angle=1B--123--12};
  \draw pic[draw,angle radius=0.8cm,"$\phi_{\mathrm{BS}}$",pic text options={shift={(-0.05,0)}}] {angle=12--2--2B};
  \draw pic[draw,angle radius=1.0cm,"$\phi_{\mathrm{MS}}$",pic text options={shift={(0.08,0)}}] {angle=2C--231--23};
  \draw pic[draw,angle radius=1cm,"$\alpha_{\mathrm{MS}}$",pic text options={shift={(-0.55,0)}}] {angle=23--3--3B};

  \node[cloud,
    draw =black,
    text=cyan,
    fill = gray!10,
    minimum width = 0.5cm,
    minimum height = 0.25cm] (c) at (4.9,-0.2) {Scatter 1};
    
  \node[cloud,
    draw =black,
    text=cyan,
    fill = gray!10,
    minimum width = 0.5cm,
    minimum height = 0.25cm] (c) at (1.8,-0.2) {Scatter 2};

\end{tikzpicture}
}
\caption{An illustration of AoAs and AoDs in the uplink channel of an IRS-aided system.}
\label{fig:irs_sys}

\vspace{-0.6cm}
\end{figure}
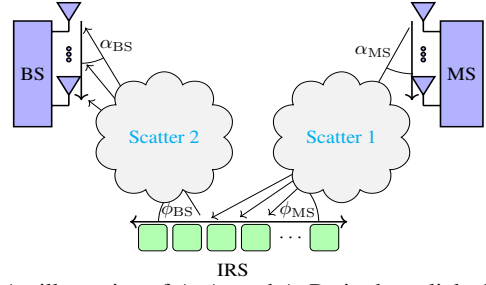

Consider an uplink MIMO millimeter-wave/terahertz band system with an MS with $M$ antennas, a BS with $B$ antennas, and a uniform linear array IRS with $L$ elements. We assume that the light-of-sight (LOS) path between the BS and MS is blocked, and the LOS paths between the BS and IRS and the IRS and MS are much stronger than the non-LOS paths. Let $\bm H_\mathrm{MS}\in \mathbb{C}^{L\times M}$ and $\bm H_\mathrm{BS} \in \mathbb{C}^{B\times L}$ denote the MS-IRS and IRS-BS channels, respectively. We assume a narrowband fading channel following the Saleh-Valenzuela model \cite{you2022structured}:
\begin{align}
\label{eq.channelmodel1}
\bm H_\mathrm{MS}&=\sum_{p=1}^{P_{\mathrm{MS}}}\sqrt{\frac{LM}{P_{\mathrm{MS}}}}\beta_{{\mathrm{MS}},p}\bm a_L(\phi_{{\mathrm{MS}},p})\bm a_{M}(\alpha_{{\mathrm{MS}}})^\mathsf{H}\\\label{eq.channelmodel2}
\bm H_\mathrm{BS}&=\sum_{p=1}^{P_{\mathrm{BS}}}\sqrt{\frac{BL}{P_{\mathrm{BS}}}}\beta_{{\mathrm{BS}},p}\bm a_{B}(\alpha_{{\mathrm{BS}},p})\bm a_L(\phi_{{\mathrm{BS}}})^\mathsf{H},
\end{align}    
where $P_{\mathrm{MS}}$ and $P_{\mathrm{BS}}$ are the number of rays. Also, for any integer $Q$ and angle $\psi$, steering vector $\bm a_Q(\psi)\in\mathbb{C}^{Q\times 1}$ is
\begin{equation}\label{eq.steer}
\bm a_Q(\psi) = \frac{1}{\sqrt{Q}}\begin{bmatrix}
1 & e^{j2\pi \delta \cos{\psi}}\cdots e^{j2\pi (Q-1) \delta \cos{\psi}}\end{bmatrix}^\mathsf{T}.
\end{equation}
 Here, we assume half-wavelength spacing, i.e., $\delta=0.5$. The angles $\phi_{{\mathrm{MS}},p}$, $\alpha_{\mathrm{MS}}$, $\alpha_{{\mathrm{BS}},p}$, and $\phi_{{\mathrm{BS}}}$ denote the $p$-th angle of arrival (AoA) of the IRS, and the angle of departure (AoD) of the MS, the $p$-th AoA of the BS, and the AoD of the IRS, respectively (see Fig.~\ref{fig:irs_sys}). Therefore, the cascaded MS-IRS-BS channel is given by $\bm H_\mathrm{BS}\diag (\bm \theta)\bm H_\mathrm{MS}$ for the IRS configuration $\bm\theta \in \mathbb{C}^{L \times 1}$. Here, the $i$-th entry of $\bm\theta$ represents the gain and phase shift due to the $i$-th IRS element. We aim to estimate the cascaded channel for any $\bm \theta$ given by $\bm H_\mathrm{BS}\diag (\bm \theta)\bm H_\mathrm{MS}$. This problem arises, for example, in beamforming problem to obtain the IRS configuration that maximizes the channel~gain~$\|\bm H_\mathrm{BS} \diag (\bm \theta) \bm H_\mathrm{MS}\|_F^2$ \cite{wang2020compressed}. However, $\vect(\bm H_\mathrm{BS}\diag (\bm \theta)\bm H_\mathrm{MS})=(\bm H_\mathrm{MS}^\mathsf{T} \odot \bm H_\mathrm{BS})\bm\theta$, where $\odot$ is the Khatri-Rao product. Therefore, the cascaded channel estimation is equivalent to estimating $\bm H_\mathrm{MS}^\mathsf{T} \odot \bm H_\mathrm{BS}$.


To estimate the channel, we send pilot symbols over $K$ time slots over which $\bm H_\mathrm{MS}$ and $\bm H_\mathrm{BS}$ are assumed to be constant. We choose $K_{\mathrm{I}}<K$ IRS configurations,
and
for each configuration, we transmit pilot data $\bm X \in \mathbb{C}^{M\times K_{\mathrm{P}}}$ over $K_{\mathrm{P}}$ time slots such that $K=K_{\mathrm{I}}K_{\mathrm{P}}$. Hence, the received signal $\bm Y_k\in \mathbb{C}^{B \times K_{\mathrm{P}}}$ corresponding to the $k$-th configuration $\bm \theta_k$~is
\begin{equation}\label{eq.basicdatamodel}
\bm Y_k=\bm H_\mathrm{BS}\diag (\bm \theta_k)\bm H_\mathrm{MS} \bm X+\bm W_k,
\end{equation}
where $\bm W_k \in \mathbb{C}^{B\times K_{\mathrm{P}}}$ is the additive white Gaussian noise with zero mean and known variance $\sigma^2$. Our objective is to estimate $\bm H_\mathrm{MS}^\mathsf{T} \odot \bm H_\mathrm{BS}$, using the data model in \eqref{eq.basicdatamodel} and the knowledge of $\bm X$ and $\{\bm Y_k,\bm \theta_k\}_{k=1}^{K_{\mathrm{I}}}$. The estimation task is challenging because of $(i)$ highly structured unknowns $\bm H_\mathrm{MS}$ and $\bm H_\mathrm{BS}$ entangled with the known quantities $\bm \theta_k$ and $\bm X$; and $(ii)$ the scaling ambiguity due to the product form $\bm H_\mathrm{BS}\diag (\bm \theta_k)\bm H_\mathrm{MS}$. In particular, if an algorithm estimates $\bm H_\mathrm{MS}$ and $\bm H_\mathrm{BS}$ separately, it cannot distinguish two solutions $(\bm H_\mathrm{MS},\bm H_\mathrm{BS})$ and  $(1/q\bm H_\mathrm{MS},q\bm H_\mathrm{BS})$, for any $q\neq0$. The following section presents our estimation algorithm.

\section{Channel Estimation Algorithms}

This section formulates the channel estimation task as a sparse recovery problem exploring angular sparsity and derives new Bayesian algorithms using the SBL framework.

\subsection{Sparse Recovery Formulation}\label{sec.sparseform}
The first step to formulate the sparse recovery problem is to disentangle $\bm\theta_k$ from the unknowns $\bm H_\mathrm{MS}$ and $\bm H_\mathrm{BS}$. For this, we vectorize both sides of~\eqref{eq.basicdatamodel} to obtain
\begin{equation}\label{eq.vecmodel1}
\bar{\bm y}_k = \big((\bm X^\mathsf{T} \bm H_\mathrm{MS}^\mathsf{T}) \odot \bm H_\mathrm{BS}\big)\bm \theta_k + \bar{\bm w}_k\in \mathbb{C}^{BK_{\mathrm{P}} \times 1},
\end{equation}
So, the received data $\bar{\bm Y}\!\in \mathbb{C}^{B K_{\mathrm{P}} \times K_{\mathrm{I}}}$ for the IRS configurations $\bm \Theta\!=\![\bm \theta_1,\ldots,\bm \theta_{K_{\mathrm{I}}}]\!\in \mathbb{C}^{L \times K_{\mathrm{I}}}$ is
\begin{equation}\label{eq.k1k2model}
\bar{\bm Y} \!=\! [\bar{\bm y}_1,\!\ldots\!,\bar{\bm y}_{K_{\mathrm{I}}}]\!=\!\big(\!(\bm X^\mathsf{T} \bm H_\mathrm{MS}^\mathsf{T}) \odot \bm H_\mathrm{BS} \big)\bm \Theta + \bar{\bm W}\! \!,    
\end{equation}
where 
 $\bar{\bm W} \!=\! [\bar{\bm w}_1,\ldots,\bar{\bm w}_{K_{\mathrm{I}}}]\! \in \mathbb{C}^{B K_{\mathrm{P}} \times K_{\mathrm{I}}}$.
Next, we leverage angular sparsity in the channel matrices $\bm H_\mathrm{MS}$ and $\bm H_\mathrm{BS}$. For this, we apply the basis expansion model by discretizing the angular domain using a set of $N$ grid angles $\{\psi_n\}_{n=1}^N$ such that $\cos(\psi_n)= 2 n/N-1$ \cite{mao2022channel}. Then, \eqref{eq.channelmodel1} and~\eqref{eq.channelmodel2} reduce to
\begin{equation}
\label{eq.channelmodelsparse2}
\bm H_\mathrm{BS}= \bm A_{B} \bm g_{\mathrm{B}} \bm g_{\mathrm{L,d}}^\mathsf{H} \bm A_{L}^\mathsf{H}  \hspace{0.3cm} \text{and }  \hspace{0.3cm}
\bm H_\mathrm{MS}= \bm A_{L}\bm g_{\mathrm{L,a}} \bm g_{\mathrm{M}}^\mathsf{H} \bm A_{M}^\mathsf{H},
\end{equation}
where for any integer $Q>0$, using \eqref{eq.steer}, we define
\begin{equation}
\bm A_{Q} = \begin{bmatrix}
\bm a_{Q}(\psi_1) & \bm a_{Q}(\psi_2)&\ldots&\bm a_{Q}(\psi_N)
\end{bmatrix}\in \mathbb{C}^{Q\times N}.
\end{equation}
Also, $\bm g_{\mathrm{B}},\bm g_{\mathrm{L,d}},\bm g_{\mathrm{L,a}},\bm g_{\mathrm{M}}\in\mathbb{C}^{N\times 1}$ are the unknown sparse channel representations. 
The non-uniform grid points in the angular domain help to reduce the computational complexity of the estimation algorithm, which is discussed in \eqref{eq.datamodelcs}. We also note that we use the same grid angle set $\{\psi_n\}_{n=1}^N$ for the AoAs and AoDs of the two channels for simplicity, but our framework and algorithm can also handle different grid sets.  


Combining \eqref{eq.channelmodelsparse2} and \eqref{eq.k1k2model}, and using the properties of the Khatri-Rao product~\cite[Lemma A1]{rao1970estimation}, we disentangle the unknown sprase vectors from the known matrices as follows:
\begin{multline}\label{eq.matrixsparseprob}
\bar{\bm Y} =\big[(\bm X^\mathsf{T} \bm A_{M}^*) \otimes \bm A_{B}\big]
	 \big[(\bm g_{\mathrm{L,a}} \bm g_{\mathrm{M}}^\mathsf{H})^\mathsf{T} \otimes (\bm g_{\mathrm{B}} \bm g_{\mathrm{L,d}}^\mathsf{H})\big] \\
	 \times (\bm A_{L}^\mathsf{T} \odot \bm A_{L}^\mathsf{H})\bm \Theta+\bar{\bm W},	
\end{multline}
where $\otimes$ denotes the Kronecker product.
Using Kronecker product's mixed-product property, we vectorize \eqref{eq.matrixsparseprob} to derive
\begin{equation}\label{eq.datamodel}
\tilde{\bm y} = (\tilde{\bm \Phi}_{\mathrm{L}} \otimes \bm \Phi_{\mathrm{M}} \otimes \bm \Phi_{\mathrm{B}}) (\bm g_{\mathrm{L,a}} \otimes \bm g_{\mathrm{L,d}} \otimes \bm g_{\mathrm{M}}^*\otimes\bm g_{\mathrm{B}}) + \tilde{\bm w},
\end{equation}
where $\tilde{\bm \Phi}_{\mathrm{L}} = \bm \Theta^\mathsf{T}(\bm A_{L}^\mathsf{T} \odot \bm A_{L}^\mathsf{H})^\mathsf{T}$, $\bm \Phi_{\mathrm{M}}=\bm X^\mathsf{T} \bm A_{M}^*$, and  $\bm \Phi_{\mathrm{B}}=\bm A_{B}$. Further, we note that the only distinct columns of $\tilde{\bm \Phi}_{\mathrm{L}}$ are its first $N$ columns~\cite{wang2020compressed}. Hence, removing the redundant columns to reduce the dimension of the representation, we get
\begin{equation}\label{eq.datamodelcs}
\tilde{\bm y} = (\bm \Phi_{\mathrm{L}} \otimes \bm \Phi_{\mathrm{M}} \otimes \bm \Phi_{\mathrm{B}})\bm g + \tilde{\bm w} = \tilde{\bm H}\bm g + \tilde{\bm w}\in\mathbb{C}^{BK\times 1},
\end{equation}
where $\bm \Phi_{\mathrm{L}}\in\mathbb{C}^{K_{\mathrm{I}}\times N}$ is the submatrix formed by the first $N$ columns of $\tilde{\bm \Phi}_{\mathrm{L}}$  and $\tilde{\bm H}=\bm \Phi_{\mathrm{L}} \otimes \bm \Phi_{\mathrm{M}} \otimes \bm \Phi_{\mathrm{B}}\in\mathbb{C}^{BK\times N^3}$. Also, we define $\bm g = \bm g_{\mathrm{L}} \otimes \bm g_{\mathrm{M}}^* \otimes \bm g_{\mathrm{B}}\in\mathbb{C}^{N^3\times 1}$ with $\bm g_{\mathrm{L}}\in\mathbb{C}^{N\times 1}$ being the scaled version of the first $N$ entries of $\bm g_{\mathrm{L,a}} \otimes \bm g_{\mathrm{L,d}}$. 
Hence, \eqref{eq.datamodelcs} translates the channel estimation problem into a sparse vector recovery problem with unknown $\bm g$. 
Using $\bm g$ and~\eqref{eq.channelmodelsparse2}, we obtain the product term in the channel as
\begin{equation}\label{eq.reconschannel}
\vect(\bm H_\mathrm{MS}^\mathsf{T} \odot \bm H_\mathrm{BS})=
(\bm \Phi_{\mathrm{A}}\otimes\bm A_{M}^* \otimes \bm \Phi_{\mathrm{B}})\bm g,
\end{equation}
where $\bm \Phi_{\mathrm{A}}$ is the first $N$ columns of $(\bm A_{L}^\mathsf{T} \odot \bm A_{L}^\mathsf{H})^\mathsf{T}$. 
Finally, the cascaded channel for a given IRS configuration $\bm\theta$ is computed as $(\bm H_\mathrm{MS}^\mathsf{T} \odot \bm H_\mathrm{BS})\bm\theta$. Thus, the rest of this section is devoted to derive an algorithm to estimate Kronecker-sparse $\bm g$ in~\eqref{eq.datamodelcs}.

\subsection{Kronecker-Sparse Bayesian Learning Algorithms}
\label{sec:sbl}


Inspired by the SBL framework~\cite{wipf2004sparse}, we impose a fictitious sparsity promoting zero-mean Gaussian prior~\cite{wipf2004sparse} (with unknown covariance) on the sparse vector $\bm g$. In our setting, to mimic the Kronecker structure, we construct the covariance matrix as $\diag(\otimes_{j=1}^3 \bm \gamma_j)$ where the vectors $\bm \gamma_1,\bm \gamma_2, \bm \gamma_3\in\mathbb{R}^{N\times 1}$ are the unknown hyperparameters corresponding to the low-dimensional sparse vectors $\bm g_{\mathrm{L}}, \bm g_{\mathrm{M}}^*$, and $\bm g_{\mathrm{B}}$, respectively. Specifically, we assume 
\begin{equation}\label{eq.sparseprior}
p(\bm g;\bm \gamma_1,\bm \gamma_2,\bm \gamma_3)=\mathcal{CN}(\bm 0,\diag(\bm \gamma)) \text{ with } \bm \gamma = \otimes_{j=1}^3 \bm \gamma_j.
\end{equation}

Then, we use Type-II maximum likelihood (ML) estimation, i.e, we first estimate the hyperparameters $\{\bm \gamma_j\}_{j=1}^3$, and using them, the estimate of $\bm g$ is the maximum point of  $ p(\bm g|\bm y;\gamma_1,\bm \gamma_2,\bm \gamma_3)$. The ML estimates of $\{\bm \gamma_j\}_{j=1}^3$ are obtained by maximizing the likelihood $p(\tilde{\bm y}; \bm \gamma_1,\bm \gamma_2,\bm \gamma_3,\sigma^2)$ with respect to them. However, this maximization problem does not admit a closed form solution, and therefore, we resort to the Expectation-Maximization (EM) algorithm~\cite{wipf2004sparse,wang2018alternative,zhang2011sparse}. The EM algorithm iterates between the E-step that provides a lower bound of the log-likelihood and the M-step which  maximizes the bound. Specifically, the $r$-th iteration of~EM~is
\begin{align}\label{eq.estep}
&\text{\bf E-step:}~ Q(\bm \gamma_1,\bm \gamma_2,\bm \gamma_3|\bm \gamma^{(r-1)})\notag\\
&\hspace{1.8cm}=\mathbb{E}_{\bm g|\tilde{\bm y};\bm \gamma^{(r-1)}}\{\log[p(\tilde{\bm y},\bm g;\bm \gamma_1,\bm \gamma_2,\bm \gamma_3)]\},\\
&\text{\bf M-step:}~
\{\bm \gamma_1^{(r)},\bm\gamma_2^{(r)},\bm\gamma_3^{(r)}\} \!=\underset{\bm \gamma_1,\bm \gamma_2,\bm \gamma_3}{\arg\max} ~Q(\bm \gamma_1,\bm \gamma_2,\bm \gamma_3|\bm \gamma^{(r-1)}),
\end{align}
where $\bm\gamma^{(r)} = \otimes_{j=1}^3 \bm \gamma_j^{(r)}$ is the $r$-th iterate. Further, we have
\begin{equation}\label{eq.post_mug}
p(\tilde{\bm y},\bm g;\bm \gamma_1,\bm \gamma_2,\bm \gamma_3)\propto p(\tilde{\bm y}|\bm g)p(\bm g;\bm \gamma_1,\bm \gamma_2, \bm \gamma_3).    
\end{equation}
Thus, from \eqref{eq.sparseprior}, the M-step can be simplified as
\begin{equation}\label{eq.mstep}
\underset{\substack{\bm\gamma_1,\bm\gamma_2,\bm\gamma_3 }}{\arg\min}\log|\diag(\bm\gamma)|+\bm d^{\mathsf{T}}\bm\gamma^{-1} \hspace{0.3cm} \text{ s.t. } 
\bm\gamma = \otimes_{j=1}^3 \bm \gamma_j,
\end{equation}
where $(\cdot)^{-1}$ is the element-wise inversion, and we define $\bm d = \diag(\bm \Sigma_{\mathrm{g}}+\bm \mu_{\mathrm{g}}\bm \mu_{\mathrm{g}}^\mathsf{H})$. Here, $\bm \mu_{\mathrm{g}}$ and $\bm \Sigma_{\mathrm{g}}$ are the mean and variance of conditional distribution $p(\bm g|\tilde{\bm y};\bm \gamma^{(r-1)})$:
\begin{equation}
\label{eq.post_meva}
\bm \mu_{\mathrm{g}} \!= \!\sigma^{-2}\bm \Sigma_{\mathrm{g}}\tilde{\bm H}^\mathsf{H}\!\tilde{\bm y},\ \bm \Sigma_{\mathrm{g}}\! =\! \left[\sigma^{-2}\tilde{\bm H}^\mathsf{H}\tilde{\bm H}\!+\!(\bm \Gamma^{(r-1)})^{-1}\!\right]^{-1},
\end{equation}
with $\bm \Gamma^{(r-1)}=\diag(\bm \gamma^{(r-1)})$.
We present two novel ways to solve~\eqref{eq.mstep}: AM-based and SVD-based, as discussed below.

\noindent \emph{AM-based Kronecker SBL (AM-KroSBL):} The AM-KroSBL solves \eqref{eq.mstep} by setting the gradient of the objective function with respect to the optimization variables to zero which gives
\begin{equation}
\label{eq.update}
    \bm \gamma_j = N^{-2}[(\otimes_{l=1}^{j-1}\bm \gamma_l^{-1})\otimes \bm I \otimes (\otimes_{l=j+1}^3\bm \gamma_l^{-1})]^\mathsf{T}\bm d,~~~ j=1,2,3.
\end{equation}
The AM-KroSBL alternatively updates $\bm \gamma_1$, $\bm \gamma_2$, and $\bm \gamma_3$ using~\eqref{eq.update} until converge. Also, to resolve the scaling ambiguity, we normalize $\bm \gamma_1$ and $\bm \gamma_2$ to have unit norm. Since the M-step is solved exactly, the algorithm inherits the convergence property of the EM algorithm~\cite{fedorov2018structured} (the details are omitted due to lack of space).
However, due to an (inner) iterative step in the M-step, AM-KroSBL is not computationally efficient. So, we next present a non-iterative method based on SVD.

\noindent \emph{SVD-based Kronecker SBL (SVD-KroSBL):} In this method, we solve \eqref{eq.mstep} without the constraint $\bm\gamma = \otimes_{j=1}^3 \bm \gamma_j$ and then project the solution to the constraint set. Specifically, we have
\begin{equation}\label{eq.mstep_svd}
\underset{\substack{\bm\gamma }}{\arg\min}\log|\diag(\bm\gamma)|+\bm d^\mathsf{T}\bm\gamma^{-1} = \bm d.
\end{equation}
To project the solution to the constraint set, 
we solve for $\bm \gamma_1,\bm \gamma_2,\bm \gamma_3$ that minimizes
$\| \bm d - \otimes_{j=1}^3 \bm \gamma_j \|$.
We further approximate this optimization problem as two rank-1 approximations solved using SVD:
\begin{equation}\label{prob.bidecom}
\underset{\substack{\bm \gamma_1,\tilde{\bm\gamma}\\\|\bm \gamma_1\|=1}}{\arg\min} \| \bm d - \vect(\tilde{\bm\gamma} \bm \gamma_1^\mathsf{T})\|,\ \underset{\substack{\bm\gamma_2,\bm\gamma_3\\\|\bm \gamma_2\|=1}}{\arg\min} \| \tilde{\bm\gamma} - \vect(\bm \gamma_3 \bm \gamma_2^\mathsf{T})\|,
\end{equation}
where we use the fact that $\vect(\tilde{\bm\gamma} \bm \gamma_1^\mathsf{T})=\bm \gamma_1 \otimes \tilde{\bm\gamma}$ and the unit-norm constraints resolve the scaling ambiguity.

\vspace{-0.3cm}
\subsection{Complexity Reduction and Analysis}
\label{sec.complexity}

SBL is known to be computationally inefficient due to matrix inversion in~\eqref{eq.post_meva}. \cite{chang2021sparse} introduced a technique to reduce the algorithm complexity using the Kronecker structure. We next present a novel technique to further improve the complexity, which can applied to both AM-KroSBL and SVD-KroSBL. Specifically, invoking the matrix inversion lemma and  the mixed product property of the Kroncker product and using the definition $\tilde{\bm H}=\bm \Phi_{\mathrm{L}} \otimes \bm \Phi_{\mathrm{M}} \otimes \bm \Phi_{\mathrm{B}}$, we rewrite \eqref{eq.post_meva}~as
\begin{multline}
\label{eq.post_varevd}
\bm \Sigma_{\mathrm{g}} =\bm \Gamma^{(r-1)} - \bm \Gamma^{(r-1)} \tilde{\bm H}^\mathsf{H}(\sigma^2\bm I + (\bm \Phi_{\mathrm{L}} \bm \Gamma_1^{(r-1)} \bm \Phi_{\mathrm{L}}^\mathsf{H})\\\otimes (\bm \Phi_{\mathrm{M}} \bm \Gamma_2^{(r-1)} \bm \Phi_{\mathrm{M}}^\mathsf{H})\otimes(\bm \Phi_{\mathrm{B}} \bm \Gamma_3^{(r-1)} \bm \Phi_{\mathrm{B}}^\mathsf{H}))^{-1}\tilde{\bm H}\bm \Gamma^{(r-1)},
\end{multline}
where $\bm \Gamma_j^{(r-1)}\!\!=\!\diag(\bm \gamma_j^{(r-1)})$, for $j=1,2,3$. 
Let the eigenvalue decomposition of the three matrices in \eqref{eq.post_varevd} be $\bm \Phi_{\mathrm{L}} \bm \Gamma_1^{(r-1)} \bm \Phi_{\mathrm{L}}^\mathsf{H} = \bm U_1 \bm\Pi_1 \bm U_1^\mathsf{H}$, $\bm \Phi_{\mathrm{M}} \bm \Gamma_2^{(r-1)} \bm \Phi_{\mathrm{M}}^\mathsf{H} = \bm U_2 \bm\Pi_2 \bm U_2^\mathsf{H}$, and $\bm \Phi_{\mathrm{B}} \bm \Gamma_3^{(r-1)} \bm \Phi_{\mathrm{B}}^\mathsf{H} = \bm U_3 \bm\Pi_3 \bm U_3^\mathsf{H}$.
Then, we derive
\begin{equation}\label{eq.newcompvar}
\bm \Sigma_{\mathrm{g}} 
= \bm \Gamma^{(r-1)}(\bm I-\tilde{\bm H}^\mathsf{H}\bm U(\sigma^2\bm I + \bm \Pi)^{-1}\bm U^\mathsf{H}\tilde{\bm H}\bm \Gamma^{(r-1)}),
\end{equation}
where $\bm U = \otimes_{j=1}^3 \bm U_j$, and $\bm \Pi = \otimes_{j=1}^3 \bm \Pi_j$. 
Combining~\eqref{eq.post_meva} and \eqref{eq.newcompvar}, the posterior mean $\bm \mu_{\mathrm{g}}$ is
\begin{multline}\label{eq.newcompmean}
\bm \mu_{\mathrm{g}} = \sigma^{-2}(\bm \Gamma^{(r-1)}\tilde{\bm H}^\mathsf{H}\tilde{\bm y} \\- \bm \Gamma^{(r-1)} \tilde{\bm H}^\mathsf{H}\bm U(\sigma^2\bm I + \bm \Pi)^{-1}\bm U^\mathsf{H}\tilde{\bm H}\bm \Gamma^{(r-1)}\tilde{\bm H}^\mathsf{H}\tilde{\bm y}).      
\end{multline}

The overall complexity of the different algorithms are summarized in Table~\ref{tab:complexity}. Here, $R_\mathrm{EM}$ is the number of EM iterations that varies across algorithms, and $R_\mathrm{AM}$ is the number of alternating iterations in~\eqref{eq.update}. We see that the proposed schemes have lower complexity compared to the KroSBL. 

In short, we present two SBL algorithms which differ from the state-of-the-art KroSBL~\cite{xu2022sparse,chang2021sparse} in two ways: the new complexity reduction technique in the E-step and the AM and SVD based-solutions in the M-step. Comparing our two schemes, AM-KroSBL enjoys the convergence guarantee while SVD-KroSBL is more efficient in practice (see Fig. \ref{fig.fig}). 


\begin{table}[htbp]
\caption{Time complexity of different versions of KroSBL}
\label{tab:complexity}
\begin{center}
\begin{tabular}{|c|c|}
\hline
\textbf{Method} & {\bf Complexity} \\
\hline
AM-KroSBL & $\mathcal{O}\big(R_\mathrm{EM}( R_\mathrm{AM} N^3 +N^3KB )\big)$\\
\hline
SVD-KroSBL & $\mathcal{O}\big(R_\mathrm{EM}( N^4 + N^3KB )\big)$ \\
\hline
KroSBL in \cite{xu2022sparse,chang2021sparse} & $\mathcal{O}\big( R_\mathrm{EM}N^6\big)$\\
\hline
\end{tabular}
\label{tab1}
\end{center}
\end{table}
\vspace{-1cm}

\begin{algorithm}[ht]
  \SetAlgoLined
  \KwData{Received signal $\tilde{\bm y}$, Pilot signal $\bm X$, IRS configuration $\bm \Theta$, noise power $\sigma^2$}
  
  
   \textbf{Parameters}: Threshold $\epsilon$ and iterations $R_{\max}$
   
  \textbf{Initialization}: $\bm \gamma_1^{(-1)} =\bm \gamma_2^{(-1)} =\bm \gamma_3^{(-1)} = \bm 1$; $\bm \mu_{\mathrm{g}}^{(0)} = \bm 0$, $\bm \mu_{\mathrm{g}}^{(-1)} = \bm 1$; and $r$ = 0.
  
   \While{$\| \bm \mu_{\mathrm{g}}^{(r)}-\bm \mu_{\mathrm{g}}^{(r-1)} \|_2 > \epsilon$ \emph{and} $r < R_{\max}$}{
  
    Define $\bm \Gamma_j^{(r-1)}=\diag(\bm \gamma_j^{(r-1)})$,$\forall j=1,2,3$
    
    Compute $\bm \Sigma_{\mathrm{g}}$ and $\bm \mu_{\mathrm{g}}$ using~\eqref{eq.newcompvar} and~\eqref{eq.newcompmean} 
    
    Obtain $\{\bm \gamma_j^{(r)}\}_{j=1}^3$ either using \eqref{eq.update} and normalizing $\{\bm \gamma_j^{(r)}\}_{j=1}^2$ (AM-KroSBL), or using \eqref{prob.bidecom} (SVD-KroSBL).

    
    
    
    Update iteration number $r \leftarrow r + 1$

  }
Compute $\bm H_\mathrm{MS}^\mathsf{T} \odot \bm H_\mathrm{BS}$ from \eqref{eq.reconschannel} with $\bm g = \bm \mu_{\mathrm{g}}^{(r)}$  
  
    \KwResult{Cascaded channel function $(\bm H_\mathrm{MS}^\mathsf{T} \odot \bm H_\mathrm{BS})\bm\theta$ }

  \caption{Our KroSBL Channel Estimation}
  \label{al.KroSBL}
\end{algorithm}

\vspace{-0.4cm}
\section{Numerical Simulation}
\label{sec:numsimu}

\begin{figure}[t]
\centering

\includegraphics[width=0.48\textwidth]{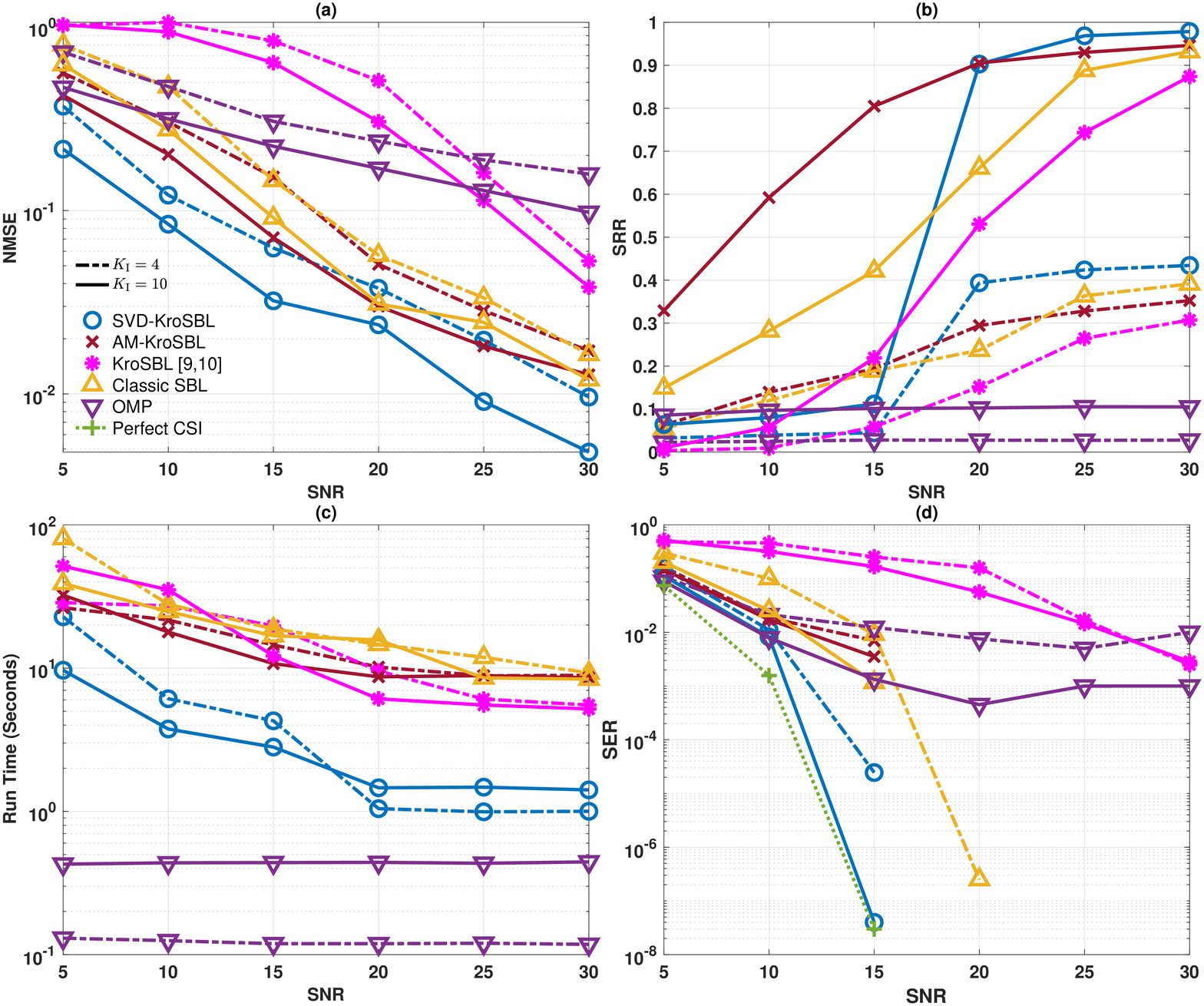}%

\captionof{figure}{Comparison of NMSE, SRR, run time, and SER of AM-KroSBL and SVD-KroSBL with the competing schemes as a function SNR when $K_{\mathrm{I}} = 4,10$, $K_{\mathrm{P}}=6$ and $N=18$.
}
\label{fig.fig}
\vspace{-0.3cm}
\end{figure}

Our simulation setting is as follows. We choose $B=16$ BS antennas, $M=6$ MS antennas, and $L=256$ IRS elements. Each entry of the IRS configurations $\{\bm \theta_k\}_{k=1}^{K_{\mathrm{I}}}$ is uniformly drawn from $\{-1/\sqrt{N}$ $,1/\sqrt{N}\}$ with $K_{\mathrm{I}}=4,10$. For each IRS configuration, we send $K_{\mathrm{P}}=6$ pilot signals. The number of grid angles is $N=18$ and all AoDs/AoAs are drawn uniformly from the grid angles. Further, the channel gains $\{\beta_{{\mathrm{BS}},p}\}_{p=1}^{P_{\mathrm{B}}}$ and $\{\beta_{{\mathrm{MS}},p}\}_{p=1}^{P_{\mathrm{M}}}$ in \eqref{eq.channelmodel1} and \eqref{eq.channelmodel2} are drawn from $\mathcal{CN}(0,1)$~ \cite{lin2021channel}. 
We use four performance metrics: normalized mean squared error (NMSE), support recovery rate (SRR), run time, and symbol error rate (SER). Here, NMSE is given as
$\frac{1}{K_{\mathrm{I}}}\sum_{k = 1}^{K_{\mathrm{I}}}\!\!\frac{\|\bm H_\mathrm{BS} \diag (\bm \theta_{k}) \bm H_\mathrm{MS} - \tilde{\bm H}_\mathrm{BS} \diag (\bm \theta_{k}) \tilde{\bm H}_\mathrm{MS}\|_F^2}{\|\bm H_\mathrm{BS} \diag (\bm \theta_{k}) \bm H_\mathrm{MS}\|_F^2},
$
with $ \tilde{\bm H}_\mathrm{BS} \diag (\bm \theta_{k}) \tilde{\bm H}_\mathrm{MS}$ being the reconstructed channel, and SRR is $ \frac{|\supp(\tilde{\bm g}) \cap \supp(\bm g) |}{|\supp(\tilde{\bm g}) - \supp(\bm g) | + |\supp(\bm g)|},
$
with $\tilde{\bm g}$ being the estimate of $\bm g$ and $\supp(\cdot)$ representing the set of indices of the nonzero entries of a vector.
We use the classic SBL~\cite{wipf2004sparse}, OMP, and the KroSBL in \cite{chang2021sparse} as benchmarks. 

From Fig.~\ref{fig.fig}, 
we observe that both SVD-KroSBL and AM-KroSBL outperform the other schemes in terms of NMSE, SRR and SER. Especially in the low SNR and low overhead (quantified by $K_{\mathrm{I}}$) regimes, our algorithms have the best NMSE.
From Fig.~\ref{fig.fig} (b), in the low SNR regime, the AM-KroSBL has the best SRR, while the SVD-KroSBL has the optimal performance in the high SNR case for both low and high overhead cases. We observe that in the low SNR regime, SVD-KroSBL outputs a sparse vector with many small terms, leading to a low SRR. But since the sparse vector is dominated by large values on the correct support, NMSE is still low. 
Further, Fig.~\ref{fig.fig} (c) indicates that SVD-KroSBL, compared with other SBL-based methods, has one order less run time. 
The high run time of AM-KroSBL is expected due to the inner loop in the M-step, yet its run time is comparable to the classic SBL and KroSBL but with better NMSE. 
Finally,  Fig.~\ref{fig.fig} (d) shows that when $K_{\mathrm{I}}=4$, our schemes possess better SER than others. In contrast, when $K_{\mathrm{I}}=10$, only SVD-KroSBL has lower SER compared to the existing schemes and approaches the oracle scheme with perfect CSI.



\section{Conclusion}
In this paper, we studied the channel estimation for IRS-aided MIMO system, exploiting the Kronecker sparse structure in the angular domain.~We presented two novel SBL-based channel estimation methods with superior performance over the state-of-the-art methods. Our AM-KroSBL enjoys strong convergence guarantees while the SVD-KroSBL is suitable for practical applications owing to its low run time.
Handling off-grid angle mismatch and extending our algorithms to multi-user case are interesting avenues for future work.


\newpage
\bibliographystyle{IEEEbib}
\bibliography{strings,refs}
\end{document}